\documentclass[proceedings]{JHEP3}

\newcommand{\sfrac}[2]{{\textstyle\frac{#1}{#2}}}

\PrHEP{PrHEP hep2001}                   
\conference{International Europhysics Conference on HEP}                

\usepackage{epsfig}                   

\title{Asymmetrically Warped Spacetimes\footnote{This talk is based
on work done in collaboration with Joshua Erlich and Christophe Grojean.}}

\author{\speaker{Csaba Cs\'aki}                       
         \\  
        Theory Division T-8, Los Alamos National Laboratory, Los Alamos, 
        NM 87545, USA\\                                          
        E-mail: \email{csaki@lanl.gov}}                       

\abstract{We investigate spacetimes in which the speed of light along flat 4D 
sections varies over the extra dimensions
due to different warp factors for the space and the time coordinates
(``asymmetrically warped'' spacetimes). The main 
property of such spaces is that while the induced metric is flat, implying
Lorentz invariant particle physics on a brane, bulk gravitational effects
will cause apparent violations of Lorentz invariance and of causality
from the brane observer's point of view. 
An important experimentally verifiable
consequence of this is that gravitational waves may travel with a 
speed different from the speed of light on the brane, and possibly 
even faster. We find the 
most general spacetimes of this sort, which are given by certain types 
of black hole spacetimes characterized by the mass and the charge of the 
black hole. We show how to satisfy the junction conditions
and analyze the properties of these space-times.} 

\begin{document}

\section{Introduction}

There is a crucial difference between the behavior of 4D and 5D theories
in the presence of non-vanishing energy densities~\cite{RubS}. 
In a 4D theory, the
presence of an ordinary 4D energy density will imply that the Universe
is expanding according to the FRW equation
\begin{equation}
H^2=\left( \frac{\dot{a}}{a}\right)^2=\frac{8\pi G_N}{3}\rho .
\end{equation}
However, in a 5D theory, a 4D energy density (a brane tension 
type object) can be balanced by a 5D energy density (a bulk cosmological
constant, since the expansion equation is of the form
$H^2 \propto \alpha \rho^2 +\beta \Lambda$, where $\rho$ is the 
4D energy density (brane tension), and $\Lambda$ is the bulk (5D) cosmological
constant. Thus, the effective 4D energy density may vanish, if the 
4D and 5D sources precisely cancel each other~\cite{RubS}. But then space is
necessarily curved, since the sources are not vanishing themselves. This
simple observation~\cite{RubS} 
is the basis of much of the recent interest in
theories with extra dimensions. A concrete realization of this idea is 
obtained by considering ``warped spacetimes'', that is spaces for which the 
form of the metric is given by
\begin{equation}
\label{warped}
ds^2=a^2(y) (dt^2-d\vec{x}^2)-dy^2,
\end{equation}
where $a(y)$ is the warp factor, and due to the form of the metric
4D Lorentz invariance is always maintained, since the induced metric
is proportional to $\eta_{\mu\nu}$. The best-known example of metrics
of this form is provided by the Randall-Sundrum (RS) model~\cite{RS}, where 
the only sources in the model are given by a brane tension $V$ and 
a bulk cosmological constant $\Lambda$, fine tuned such that the 
effective 4D cosmological constant vanishes. This model is extremely 
interesting, since it may localize gravity to the brane with a positive
tension, and thus avoid the need for compactification of the extra 
dimension. A slightly modified version of the model where the fifth dimension
ends on a second brane (with negative tension) may solve the hierarchy problem
due to the exponential falloff in the strength of gravity. A second 
interesting example of a metric of the form (\ref{warped}) is 
provided by the ``self-tuning'' brane models of \cite{selftune}.
In these models, there is an extra bulk scalar field $\Phi$ besides the
sources of the RS model, which make it possible for a flat solution 
to exist for any values values of the brane tension (while no other 
maximally symmetric solutions exist). However, these solutions are nakedly
singular at a finite distance from the brane, and the effective 
cosmological constant vanishes only after fine-tuning is reintroduced 
into the theory~\cite{Nilles}.  

\section{Asymmetric Warping}
Here we will consider a slightly modified version of the metric,
which takes the 
form \cite{CEG}
\begin{equation}
\label{asymm}
ds^2=a^2(y)dt^2-  b^2(y)d\vec{x}^2 - c^2(y) dy^2,
\end{equation}
where the warp factor for the space and time components of the metric differ,
$a\neq b$.
Some of the properties of such space-times
were also investigated in \cite{asymm}.

The first thing that one learns from the metric in (\ref{asymm}) is that the
induced metric at any point $y=y_0$ is still flat, however one needs
to use a different ($y$-dependent) rescaling of the time coordinate $t$ 
to obtain a flat induced metric $\eta_{\mu\nu}$. Thus we expect that in 
these metrics 4D Lorentz invariance would be locally respected, but broken 
globally. Therefore, if the standard model particles were to be localized
at a fix value of $y$, then particle physics would appear to be completely
Lorentz invariant, still leaving open the possibility that gravitational 
effects violate Lorentz invariance. What these effects could be is most easily
seen by considering the propagation of massless fields in such a space-time.
The local speed of light is given by $a(y)/b(y)$, which by definition
is varying along the extra dimension. This is analogous to 
propagation of light in a medium with a changing index of refraction,
governed by Fermat's principle. Thus if the local speed of light increases
away from the brane, then it might be advantageous for the gravitational
waves to bend into the extra dimension, and arrive earlier than
electromagnetic waves which are forced to travel with the local speed of light
(see Fig.~\ref{fig:brane}). This also implies, 
that regions that one thought not
to have been in causal contact with each other may have been in contact 
after all, leading to apparent violation of causality. We have to 
stress that this is only an apparent violation from the brane observer's
point of view, and the full 5D theory is completely causal (there is no 
propagation backwards in time, that is there are no closed time-like curves).
It also implies that these theories may have an unconventional 4D effective
theory, which may circumvent the no-go theorem of Weinberg for the 
adjustment of the cosmological constant.

\section{Solutions with Asymmetric Warping}
Here we find the solutions of the form (\ref{asymm}) corresponding to the
simplest possible physical sources in the bulk and on a brane, assumed 
to have a $Z_2$ symmetry (for details of this symmetry and more details 
of the solution see \cite{CEG}). We assume that the metric is homogeneous and
isotropic along the brane, thus taking the form 
\begin{equation}
\label{generalmetric}
ds^2= -n^2(t, r) \,dt^2
+ a^2(t,r) d\Sigma_k^2
+ b^2(t,r) \,dr^2,
\end{equation}
where 
$d\Sigma_k^2= d\sigma^2/ (1-k l^{-2} \sigma^2) +\sigma^2 d\Omega_2^2$
is the metric of the spatial 3-sections, with curvature parameter $k$,
$l$ being a parameter with dimension of length that will be
set to the length scale given by the cosmological constant in the bulk.
Using Birkhoff's theorem, this solution 
can always be transformed into the following simple form:
\begin{equation}
	\label{eq:metric}
ds^2=
-h(r)\,dt^2
+l^{-2} r^2\, d\Sigma_k^2
+h(r)^{-1}\,dr^2,
\end{equation}
which describes a black hole space-time. Depending on the sources 
in the problem this may be an ordinary black hole, an AdS black hole, a 
Reissner-Nordstrom (RN) black-hole, an AdS-RN black hole, etc. The most 
important cases for us are the AdS black hole solution for which
\begin{equation}
	\label{eq:AdSSch}
h(r)=k +\frac{r^2}{l^2}-\frac{\mu}{r^2},
\ \ \ l^{-2}=-\sfrac{1}{6} \kappa_5^2 \Lambda_{bk},
\end{equation}
and the AdS-RN solution in which case
\begin{equation}
	\label{eq:AdSRN}
h(r)=k + \frac{r^2}{l^2} - \frac{\mu}{r^2} + \frac{Q^2}{r^4},
\end{equation}
where $\mu$ is the mass and (for the RN case) 
$Q$ the charge of the black hole. Thus we know what the form of the 
solutions in the bulk can be, but in order to find a viable metric,
a brane has to be introduced and the Israel junction conditions at the 
brane have to be satisfied. The details of the matching procedure are 
described in \cite{CEG}, here we summarize the most important features.
For the case of an AdS BH solution, there is only one new parameter,
the mass of the black hole $\mu$, and the junction conditions will still
require a fine tuning between the energy density on the brane and the 
bulk cosmological constant. The black hole singularity is naked if in the 
equation of state on the brane $p=\omega \rho$ the parameter $\omega \geq -1$.
For the case of the AdS RN BH solution there are two new parameters
$\mu$ and $Q$, and if the $Z_2$ parities are chosen appropriately 
no fine tuning will be needed to satisfy the junction conditions. The
naked BH singularity is again only avoidable for  $\omega \leq -1$ \cite{CF}.
Once we obtain a solution with a horizon, one can show that the space-time
can be cut at the horizon without reintroducing fine-tuning into the model.
One can also show \cite{CEG}, that except for generic values of the 
parameters of the theory the only maximally symmetric solution on the brane
is the flat solution, thus there is no AdS and dS solution on the brane
for this setup.

\section{Lorentz Violations and Experimental Signatures}
\begin{figure*}[tb]
\centerline{\includegraphics*[bb=170 2 450 788,angle=-90,width=14cm]{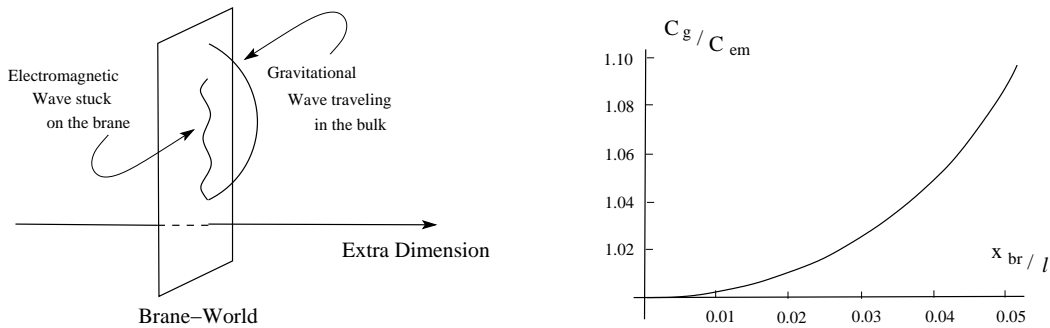}}
\caption[]{A graviton emitted on the brane will travel along
a geodesic in the bulk before returning to the brane.
A photon emitted at the same time can propagate only along the brane and may
wander a shorter distance along the brane than the graviton in the same time.
The 4D effective propagation speed of gravity
is distance dependent ($x_{br}$ is the distance traveled along the brane
and $l$ characterizes the curvature of the bulk).
}
\label{fig:brane}
\end{figure*}

Once the possible solutions with asymmetrically warped spaces are found,
we can ask whether any of the phenomena involving violations of causality
would occur in these solutions. For the general space-time of the form
(\ref{eq:metric}) the local speed of light is given by 
$c^2(r)=h(r) l^2/r^2$. One would need a growing $c(r)$ away from the brane.
Examining the full solutions that also incorporate the junction conditions
we find that this indeed may happen for a significant 
fraction of the parameter space. An example for such a case is
illustrated in Fig. \ref{fig:brane}. 

Finally we address the issue whether any of these effects could be 
experimentally observable. The simplest evidence for asymmetrically 
warped spaces would be to measure the speed of gravitational waves and find 
that it is larger than the speed of light in the vacuum. 
The LIGO experiment
may be able to detect gravitational waves from type II supernovae up to a 
distance of about 20 Mpc 
($\sim 6\cdot 10^7$ ly). For objects of such a
distance even a tiny difference in the speeds of gravitational and 
electromagnetic waves would cause a huge time difference, and thus the 
possible values of $\mu$ and $Q$ could be severly constrained.
In fact, the limitations of such measurements are likely not to lie in the
time resolution of the gravitational
and the light signal, but rather the opposite problem: if there is an
appreciable difference in the propagation speeds then due to the huge
distance to the expected sources the arrival time differences could turn
out to be way too big to be able to identify the fact that the source
for the gravitational wave and the light was the same. For a supernova 20
Mpc away from us, and very conservatively assuming that the arrival time
difference should be less than 5 years, in order to be able to actually
detect the different arrival times one needs to have the difference in
the speeds to be less than $\frac{\delta c}{c} \leq 10^{-7}$. Otherwise
the gravitational wave experiments will simply not be able to 
identify the source for the observed gravitational waves. Type I 
supernovae could likely be detected by LIGO only if they happen within our
galaxy. These are very rare, however assuming the best possible scenario
one could see a supernova a few hundred thousand light years from us. In this
case (again assuming a very conservative time difference of 5 years) 
the maximum value of $\frac{\delta c}{c}$ that could be tested is 
of the order of $10^{-3}$.

\section{Conclusions}
We have considered space-times where the local speed 
of light varies along the extra dimensions. Such asymmetrically warped
spaces could have interesting physical properties, since gravitational
waves might propagate faster than electromagnetic waves. This could
be verified experimentally by the upcoming gravitational wave experiments.
We have found a large class of such solutions in forms of various 
black hole spacetimes characterized by the mass and the charge of the 
black hole, showed how the junction conditions can be satisfied, and 
analyzed the physical properties of these solutions.

\vspace*{-0.2cm}
\section*{Acknowledgements}
{\small This talk was based on~\cite{CEG} which was done in collaboration with 
Joshua Erlich and Chris\-tophe Grojean. The author is an Oppenheimer 
Fellow at the Los Alamos National Laboratory, supported by the 
US Department of Energy under contract W-7405-ENG-36, and in part by a 
DOE OJI grant.}



\end{document}